# A Letter to the NSF Astronomy Portfolio Review: LSST is Not "Big Data"

## David Schlegel (Lawrence Berkeley National Lab)
## 31 January 2012

LSST promises to be the largest optical imaging survey of the sky. If we were fortunate enough to have the equivalent of LSST today, it would represent a "fire hose" of data that would be difficult to store, transfer, and analyze with available compute resources.

LSST parallels the SDSS compute task which was ambitious yet tractable. By almost any measure relative to computers that will be available (thanks to the steady progression of Moore's Law), LSST will be a small data set. LSST will never fill more than 22 hard drives. Individual investigators will be able to maintain their own data copies to analyze as they choose.

NSF and DOE resources for data reduction should reflect the scale of the problem. A centralized computing model is not necessary, and could introduce a barrier to science. An over-investment in LSST survey data reduction should not squeeze out other survey opportunities on the existing NOAO 4-m telescopes.

**Data Storage**

The transistor count of CPUs and the pixel count of telescope focal planes have advanced at essentially the same pace, doubling every 2 years as predicted by Moore's Law (Fig 1). Between 1985 and 2010, the transistor count (and raw compute speed) of CPUs has increased by a factor of 5000, from the Intel 80386 to the 8-core Xeon Nahalem. Telescope focal planes have increased by the same factor, from the single-CCD focal planes with the Tektronix $512^2$ to the PanSTARRS focal plane with a 1.4 giga-pixel mosaic.

Spinning, hard drive storage has increased at a rate even faster than Moore's Law. The capacity of a $400 commercial drive has doubled every 1.6 years on average, reaching a capacity of 3 TB per drive today. At the same time, the technology has been developed for easily distributing data across many drives: RAID storage allowed the transparent use of many disks within a single enclosure, and distributed file systems (such as Lustre) will allow even larger data volumes with a single mount point.

The SDSS imaging survey pushed the limits of a data volume that could be spinning. When it began science operations in September 1998, its camera would fill the largest 8 GB hard drives in 25 minutes. With the data set increasing linearly with time, and storage doubling every 1.6 years, the peak storage cost occurs 2.3 years into any survey. For SDSS, if one were to re-purchase disk at appropriate intervals, the peak storage cost was January 2002, when 106 hard drives would be necessary to store the compressed 3.2 TB of imaging. These data were put on spinning disk at Princeton at that time, with the added complication of requiring over 100 disk mount points. Today we can store the entire compressed 30 TB data set on ten disks, and many institutions have chosen to do just that.

The LSST storage is much easier than SDSS when normalized to available technology. If LSST begins routine operations on Jan 1, 2021, it will fill a then-available 120 TB hard drive every 25 nights. The peak storage cost is predicted to occur in March 2023 when 22 disks will be needed. The completed 10-year survey is reasonably predicted to fit on 3.5 disks if purchased in 2031.) Any institution (or individual!) will be able to maintain a spinning copy of the entire LSST data set with an investment of $15k per year or less.

**Compute Time**

LSST would represent a comparable compute challenge to SDSS if it were to enter routine operations today. The focal plane is larger by a factor of 25 and a standard computer is faster by that same factor (c.f. Fig 1). By the time LSST enters routine operations, computers are predicted to be faster by another factor of 30 making the data reduction less challenging by that same factor.

The SDSS and LSST imaging surveys are "embarrassingly parallel" computing challenges (to use the technical term). For the LSST key science projects, each pointing on the sky need only analyze other pointings in the same location.

A limited number of tasks need knowledge from other pointings. These include global calibrations of the photometry and astrometry. Such problems have been solved in the past, for example with the photometric recalibration of SDSS done by Padmanabhan et al 2008 (http://arxiv.org/abs/astro-ph/0703454).

**Will There Ever Be a Large Survey?**

This discussion begs the question of whether there will ever be an astronomical survey that is large relative to compute or storage capabilities. For optical astronomy, the answer may indeed be "no."

One could imagine the perfect optical instrument as one that could time-tag and energy-tag every photon hitting a focal plane. The dark night sky integrates to 0.055 photons/sec/$cm^2$/$arcsec^2$ over the wavelength range 3500-10,000 Ang. As most of these photons are from the OH emission lines in the near-IR, the bright (moonlit) sky is not very different. If the entire 9.6 $deg^2$ of the LSST focal plane could be outfitted with a 100% efficient detector, this would collect $6.9 \times 10^{16}$ photons per night. If 8 bytes per photon were required to encode the position, energy and time information of each photon, this integrates to 550 PB / night.

This perfect instrument would represent a data challenge comparable to SDSS should such an instrument become available before late 2033. Since no such detector is in the foreseeable future, such musings are probably not relevant for the current NSF-AST Portfolio Review.

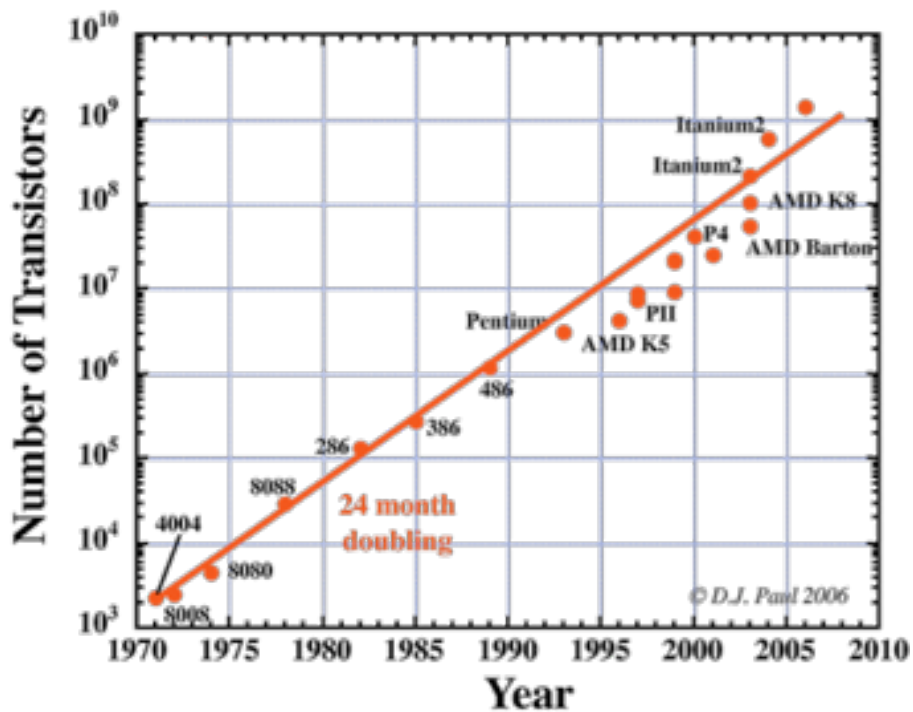

Fig 1: The processing speed of computers has followed the prediction of Moore's Law, doubling every 2 years (http://www.sp.phy.cam.ac.uk/~SiGe/Moore%27s%20Law.html).

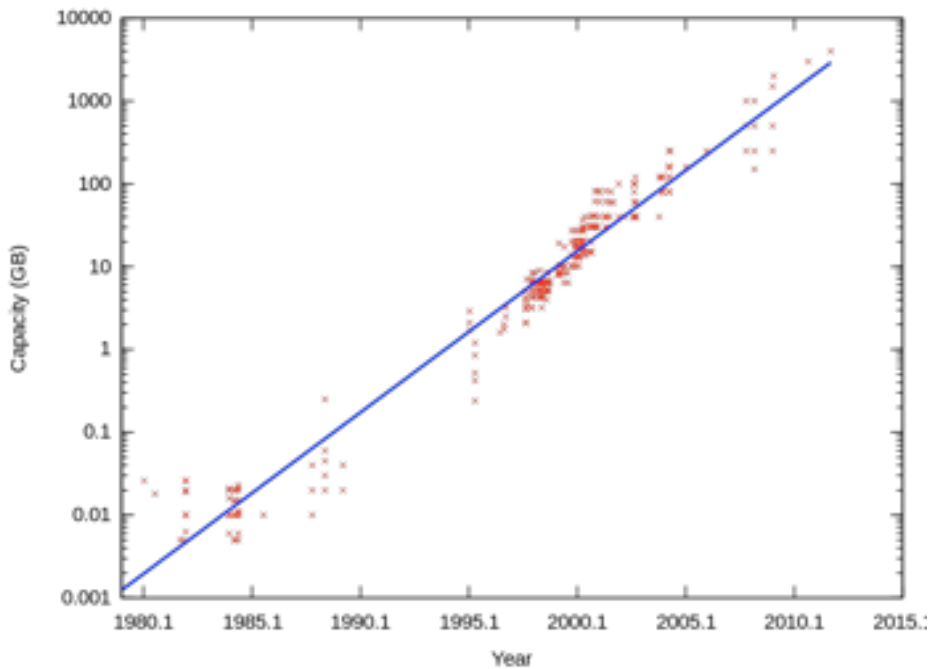

Fig 2: Storage on hard drives has increased faster than Moore's Law, doubling every 1.6 years since 1980 (http://en.wikipedia.org/wiki/Hard_drive).